\documentclass[aps,prl,preprint,superscriptaddress]{revtex4-1}
\usepackage[utf8]{inputenc}
\usepackage{amsmath,amssymb,bbm,color,lmodern,graphicx}
\usepackage[section]{placeins}
\newcommand{\mathe}{\mathrm{e}}
\begin{document}
%===================================================================== 
%===================================================================== 
\title{Hamiltonian dynamics of the SIS epidemic model with stochastic
  fluctuations}    
\author{Gilberto M. Nakamura}
\email{gmnakamura@usp.br}
\affiliation{Faculdade de Filosofia, Ci\^{e}ncias e Letras de
  Ribeir\~{a}o Preto (FFCLRP), Universidade de S\~{a}o Paulo,
  Avenida Bandeirantes 3900, 
  14040-901 Ribeir\~{a}o Preto, Brazil}  
\affiliation{Instituto Nacional de Ci\^{e}ncia e Tecnologia -
  Sistemas Complexos (INCT-SC)}
\author{Alexandre S. Martinez}
\email{asmartinez@usp.br}
\affiliation{Faculdade de Filosofia, Ci\^{e}ncias e Letras de
  Ribeir\~{a}o Preto (FFCLRP), Universidade de S\~{a}o Paulo,
  Avenida Bandeirantes 3900, 
  14040-901 Ribeir\~{a}o Preto, Brazil}  
\affiliation{Instituto Nacional de Ci\^{e}ncia e Tecnologia -
  Sistemas Complexos (INCT-SC)}

%=====================================================================
%===================================================================== 
\begin{abstract}
  Empirical records of epidemics reveal that fluctuations are
  important factors for the spread and prevalence of infectious
  diseases. The exact manner in which fluctuations affect spreading
  dynamics remains poorly known. Recent analytical and numerical
  studies have demonstrated that improved differential equations for mean and
  variance of infected individuals reproduce certain regimes of the
  SIS epidemic model. Here, we show they form a
  dynamical system that follows Hamilton's equations, which allow us
  to understand the role of fluctuations and their effects on
  epidemics. Our findings show the Hamiltonian is a constant of motion
  for large population sizes. For small populations, finite size
  effects break the temporal symmetry and induce a power-law decay of
  the Hamiltonian near the   outbreak onset, with a parameter-free
  exponent. Away from onset, the Hamiltonian decays exponentially
  according to a constant relaxation time, which we propose as an
  indicator of the strength of the epidemic when fluctuations cannot
  be neglected. 
  %% We propose its use to
  %% measure the strength of epidemics when fluctuations cannot be
  %% neglected.  
\end{abstract}
%=====================================================================
%===================================================================== 
\maketitle
%=====================================================================
%===================================================================== 

Models of disease transmission, or epidemic models for
short, have been an integral part of the epidemiological toolkit,
dating back from pioneer models of Kermack and McKendrick
\cite{kermackProcRSocA1927}. The main goal of epidemic models can be  
summarized as the ability to accurately predict spreading patterns
of a given communicable disease afflicting a specific population.
These models allow decision makers to assess the various intervention
strategies available to them and to plan accordingly. Several
approaches have been developed to model disease outbreaks
\cite{heesterbeekScience2015}, namely, compartmental equations,
stochastic equations, agent-based simulations, etc. Each approach
suits a particular aspect of the outbreak being studied, built upon 
hypotheses compatible with empirical records or based on a
phenomenological context. They include, but are not restricted to,
biological content of the disease, mechanisms behind pathogen
transmission, social interactions among the target population and its
spatial structure \cite{satorrasRevModPhys2015}. By the same
token, different models for the same disease and population may
produce inconsistent results, possibly due to conflicting underlying
hypotheses. For instance, the random-mixing hypothesis -- i.e. every
element in the population has an equal change to interact with any
other element -- seems reasonable to model pathogen transmission for
airborne disease like influenza, but it seems equivocated for sexually
transmitted diseases \cite{keelingJRSoc2005}.  

Despite the significant advances obtained in the past few decades,
several challenges still remain open. One issue concerns the
failure to account effects unrelated to diseases themselves, such as
vaccination skepticism, which ultimately reduces children immunization
rate. Outbreaks of treatable communicable disease, like measles, are
on the rise \cite{hornePNAS2015}. Another issue deals with
understanding the complex dynamics and processes behind infections in
both small and large scales
\cite{pellisEpidemics2015,brittonEpidemics2015,bansalEpidemics2015}.
To put it simply, there are too many variables and their effects are not 
entirely known due to the non-linear nature of the problem. As a
consequence, the full extent of variable changes or their fluctuations
remains poorly understood, which may produce sub-optimal intervention
strategies. As an example, detailed field data from recent Ebola
epidemic have shown that smaller outbreaks from different localities
are asynchronous \cite{bansalPhysLifeRev2016}. The lack of
synchronization between different populations reduces the likelihood
of pathogen eradication on a global scale, as long as migration is
allowed in some form \cite{ruxtonJRSocInterface1994,earnProcRSocB1998}.
The effects of migration and  spatial structures in epidemic models
and pathogen variability have been under investigation for some time
\cite{biekJRSoc2007,hastingsEcol2010}, and they have been linked 
to chaotic dynamics in local population \cite{allenNature1993}.
Experiments on the effects of migration between metapopulations,
i.e. similar populations but spatially separated, subjected to
temporal fluctuations have shown that pathogen prevalence is greatly
influenced by the nature of the fluctuation \cite{duncanProcRSocB2013},
highlighting the interplay between synchronization and pathogen
prevalence in epidemics \cite{parisiJRSocInterface2016}.

Traditionally, the detailed examination of fluctuations -- either
temporal or spatial -- and their effects on system dynamics have been
largely described by correlation functions
\cite{duncanProcRSocB2013,cardy}. More recently, autocorrelation functions 
have been used to reveal the nature and general aspects of
fluctuations in a simple agent-based epidemic models for a population
of size $N$, in which temporal fluctuations are divided into two broad
classes: gaussian and non-gaussian \cite{nakamuraArxiv2018}. In the
gaussian regime, the prevalence of the disease is well described by
its instantaneous average, finite variance, and higher cumulants can
be neglected. This is remarkable as it allows 
one to derive the exact contributions of fluctuations to disease
outbreaks in the asymptotic limit $N \gg 1$. Here, we show that the
dynamical equations obtained in Ref.~\cite{nakamuraArxiv2018} form a
Hamiltonian dynamical system, and the way external noise can be
incorporated to model disease outbreaks. This approach allows us to
discuss quantitatively the relevant scales of the problem, and
interpret the resulting Lagrangian and canonical transformations.
%% The
%% paper is organized as follows. A brief review of the SIS model is
%% presented in Sec.~\ref{sec:model}. Sec.~\ref{sec:equations} outlines
%% the procedure to derive the dynamical system with intrinsic noise. The
%% Hamiltonian  function is obtained in Sec.~\ref{sec:hamilton}.
%% The Lagragian is shown to be twice the standard deviations, and the
%% applicability of canonical transformation is presented in
%% Sec.~\ref{sec:lagrange} to show the consistency of our
%% approach. Conclusions and final remarks are disclosed  
%% in Sec.~\ref{sec:conclusion}.

\begin{figure}
  \includegraphics[width=0.95\columnwidth]{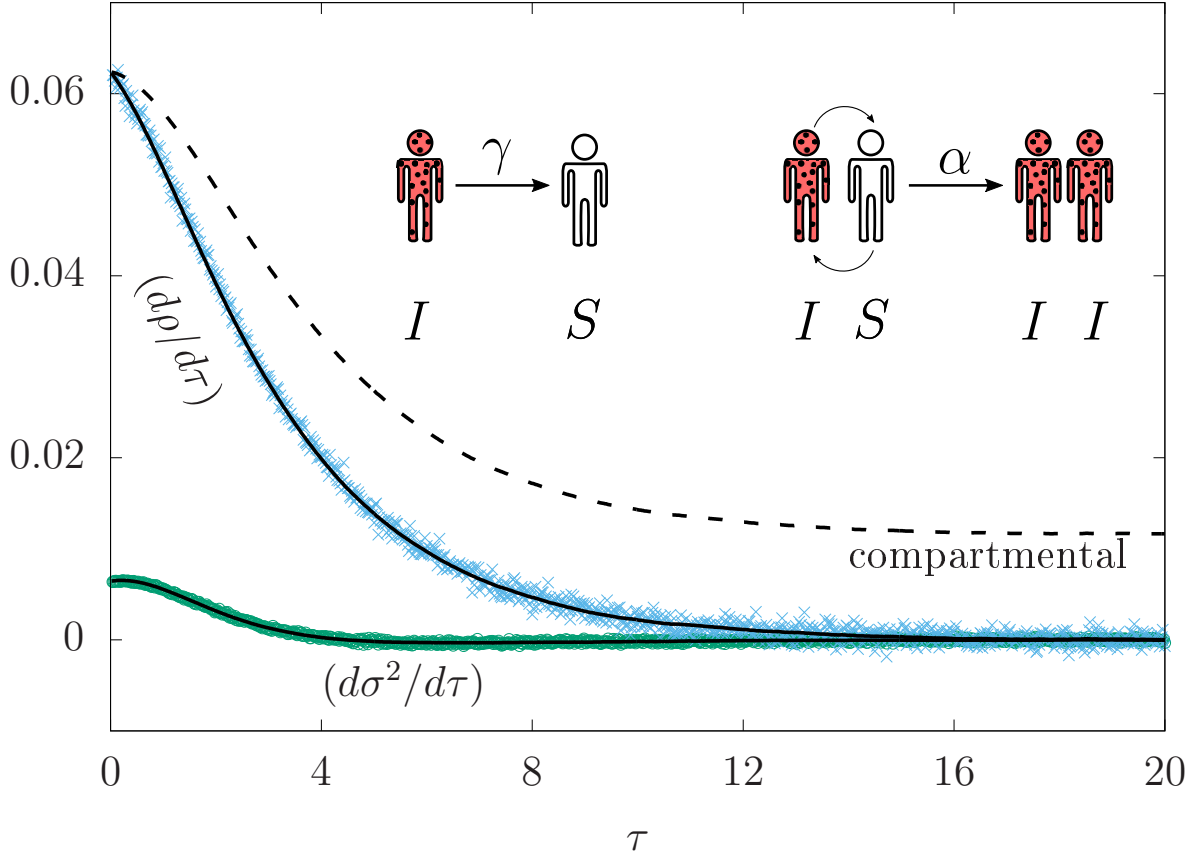}
  \caption{\label{fig:fig1} Numerical simulations of the SIS model.
    (inset) Infected hosts (I) recover to susceptible state (S) with
    rate $\gamma$ (left). Adequate interaction between an infected
    host with a susceptible one may trigger a new infection with rate
    $\alpha$ (right).
    Stochastic effects are far more prominent for small
    population sizes ($N=50$, $\gamma/\alpha = 1/2$ ), reducing the accuracy of compartmental
    equations. The time derivative of the density of infected $\rho$
    extracted directly from data (cross) using forward-derivative
    agrees with the RHS of Eq.~(\ref{eq:imp1}), as a function of the
    density and variance. The dashed line shows the expected RHS of
    compartmental equation Eq.~(\ref{eq:compartmental}). The equation of
    motion (line) for $d\sigma^2/d\tau$ in Eq.~(\ref{eq:imp2}) also agrees with
    simulated data (circles).
  }
\end{figure}

%% \section{Model}
%% \label{sec:model}
\textit{Model.}
We begin our discussion using the susceptible-infected-susceptible
(SIS) epidemic model. The SIS model describes the dissemination of a
single communicable disease in a susceptible population of size $N$.
The transmission of the pathogen occurs when infected hosts transmit the
disease pathogen to healthy susceptible individuals. The infectious
period extends throughout the whole course of the disease until
recovery of the patient, warranting a two-stage model: either infected
or susceptible. The essence of the model is summarized by inset in
Fig.~\ref{fig:fig1}. 
%% \begin{figure}
%%   \includegraphics[width=0.75\columnwidth]{fig1.eps}
%%   \caption{\label{fig:fig1} SIS model.  Infected hosts (I) recover to
%%     susceptible state (S) with rate $\gamma$ (left). Adequate
%%     interaction between an infected host with a susceptible one may 
%%     trigger a new infection with rate $\alpha$ (right). }
%% \end{figure}

The traditional formulation of the problem assumes the random-mixing
hypothesis (see Introduction) holds for a large population size $N
\gg 1$, compromised of statistically equivalent individuals. Under
these circumstances, the only relevant variable is the instantaneous
density of infected elements $\rho(t)$, which means that fluctuations
can be safely neglected. Furthermore, $\rho(t)$ decreases with rate
$\gamma\,\rho$, where $\gamma$ is the recovery rate. New infections
per unit of time (disease incidence) are proportional to $\alpha
\rho(1-\rho)$, i.e., they depend on the chance that infected elements
interact with susceptible ones, with intensity given by the
transmission rate $\alpha$. This picture provides an interpretation
where $\rho(t)$ is continuously exchanged  between two 
compartments, leading a simple description called compartmental
equation: $   d\rho(t)/dt = \alpha \rho (1 -\rho)  -
  \gamma \rho $.
%% \begin{equation}
%%   \frac{d}{dt} \rho(t) = \alpha \rho (1 -\rho)  -
%%   \gamma \rho.
%% \end{equation}
For the sake of convenience, redefine the timescale as $\tau \equiv
\alpha t$  and $\rho_0 \equiv 1 - \gamma/\alpha$, so that 
\begin{equation}
  \label{eq:compartmental}
  \frac{d}{d \tau}\rho(\tau) = \rho (\rho_0 -\rho).
\end{equation}
Clearly, the equilibrium density can either be $\rho_{\textrm{eq}} =0$
or $\rho_{\textrm{eq}} = \rho_0$. Also, $\rho_0$ is related to the
basic reproduction number $R_0 = N (\alpha/\gamma) $ which provides an
estimate on the number of new infections per generation
\cite{caliriJBioPhys2003}. 

In light of its long age, compartmental equations have met
considerable success in predicting the time evolution of disease
outbreaks, providing valuable insights for intervention strategies and
funding allocation \cite{murray}. However, outbreaks that fail to meet
the underlying hypotheses (random mixing and large population of
statistically equivalent elements) can contradict compartmental
equations. These inconsistencies are largely attributed to stochastic
effects and their inherent fluctuations \cite{heesterbeekScience2015}.

%=====================================================================
%===================================================================== 
%% \section{Improved compartmental equations}
%% \label{sec:equations}
\textit{Improved compartmental equations.}
Stochastic variables are known to cause the emergence of critical
phenomena in computer simulations of epidemic models, under certain 
parameter ranges
\cite{rhodesProcRSocB1997,rhodesTheorPopulBio1997}. One key ingredient
common to almost every critical phenomena points is the scale
invariance of fluctuations
\cite{chialvoPhysRevLett2017,stanleyRevModPhys1999}. This special
symmetry remains the foundation of cooperative phenomena and critical
phase transitions, whose contributions spans over a broad set of
research fields such as condensed-matter, quantum field theories, and
neuroscience to name a few
\cite{nakamuraJPhysA2010,wilsonPhysRevD1974,kogutRevModPhys1979,chialvoPhysRevLett2005,bialekNature2006}. In 
these special systems, fluctuations of descriptive variables occur in
all sizes and, ultimately, dictate the general behavior of the
problem. It thus begs the question: if critical  behavior has been
observed previously in disease outbreaks \cite{rhodesProcRSocB1997},
why fluctuations have been neglected in the mathematical modeling of
epidemics?

So far, the effects of stochastic fluctuations on general epidemics
remains poorly known. New experiments on this subject provide evidence
that temporal fluctuations can drastically alter the prevalence of
pathogens \cite{duncanProcRSocB2013}. Spatial heterogeneity also
introduces an extra layer of complexity as it may trap or delay the
pathogen transmission \cite{biekJRSoc2007}. As a result, requirements
of statistical equivalence may not hold for all scales. To deal with
this issue, stochastic formulations and numerical simulations have
been the default tools to investigate fluctuations in disease
outbreaks.
%% In what follows, we outline the general ideas 
%% behind the agent-based approach to describe the SIS model. A more
%% detailed account of the results listed here are explained in detail in
%% Refs.~\cite{nakamuraSciRep2017,nakamuraArxiv2018}. 

Our discussion assumes the disease spreading follows a Markov chain in
discrete time  $\delta t$. Moreover, $\delta t$ is such that at most
a single recovery or transmission event is likely to occur during the
course of its duration. Under these requirements, the master equation
of the SIS model in discrete time reads
\begin{equation}
  \label{eq:master}
  \frac{d P_{\mu}(t)}{dt} = - \sum_{\nu=0}^{2^N-1}H_{\mu\nu}P_{\nu}(t).
\end{equation}
Here, $P_{\mu}(t)$ refers to the instantaneous probability to observe
the system in the $\mu$-th configuration. Configuration labels follow
the binary ruling $\mu = n_0 2^0+ n_1 2^1+ \cdots + n_{N-1} 2^{N-1}$,
where $n_k = 1$ if the $k$-th agent is infected, or $n_k=0$ otherwise,
with $k=0,1,\ldots,N-1$. For instance, for $N=3$, the configuration
$\lvert \mu = 3 \rangle = \lvert 110 \rangle$ states that only the
agent with label $k=2$ is susceptible. The matrix elements $H_{\mu\nu}$
express the transition rates from configuration $\nu$ to
configuration $\mu$. By virtue of probability conservation, in each
time step the transition rules satisfy $\sum_{\mu}H_{\mu\nu} =0$. The
matrix elements $H_{\mu\nu} = \langle \mu | \hat{H} | \nu\rangle$ are
computed from projections on the time step operator
\begin{equation}
  \hat{H} =
  \frac{\alpha}{N}\sum_{k,\ell=0}^{N-1}A_{k\ell}(1
  -\hat{n}_k-\hat{\sigma}^+_k)\hat{n}_{\ell}+
  \gamma\sum_{k=0}^{N-1}(\hat{n}_k-\hat{\sigma}^-_k),
\end{equation}
where $A_{k\ell}$ is the adjacency matrix, $\hat{n}_k$ represents the $k$-th
occupation operator (with eigenvalues $n_k=1$ if infected, $0$
otherwise), and $\hat{\sigma}_k^{+}$ are the localized spin-$1/2$
ladder operators that produce the transition $S\rightarrow 
I$. Clearly, $\hat{\sigma}_k^{-}$ produce the opposite transitions,
$I\rightarrow S$ in relation the $k$-th agent. As notation, the hat
symbol always accompanies operators to quickly distinguish them from
numbers. 

The master equation Eq.~(\ref{eq:master}) provides the means to
evaluate the time evolution of relevant statistical moments of
$\rho(t)$. Notice that the average density of infected agents in the
system reads $ \langle \rho(t)\rangle = (1/N)\sum_{\mu=0}^{2^N-1}
\sum_{k=0}^{N-1} \langle \mu \rvert  \hat{n}_k \lvert \mu \rangle
P_{\mu}(t)$. 
%% \begin{equation}
%%   \langle \rho(t)\rangle = \frac{1}{N}\sum_{\mu=0}^{2^N-1}
%%   \sum_{k=0}^{N-1} \langle \mu \rvert  \hat{n}_k \lvert \mu \rangle P_{\mu}(t).
%% \end{equation}
Applying the time derivative, and using Eq.~(\ref{eq:master}), one arrives
at the equation of motion for $\langle \rho(t) \rangle$. Useful
expressions are known only for a few types of adjacency matrix
$A$. The simplest ones are proportional to $A_{k\ell} =
1-\delta_{k\ell}$, which recovers the random mixing hypothesis. In
those particular instances, the complete time evolution of the system
comprehends a set of hierarchical equations that involves the
statistical moments of $\rho(t)$, as shown in
Ref.~\cite{nakamuraSciRep2017}. More explicitly
\cite{nakamuraArxiv2018}, the first two equations for instantaneous
mean $\langle \rho \rangle$ and variance $\sigma^2 = \langle
\rho^2\rangle - \langle \rho \rangle^2$ are
%\begin{widetext}
\begin{subequations}
  \label{eq:imp}
  \begin{align}
    \label{eq:imp1}
    \frac{d\langle \rho \rangle}{d \tau} &= \langle \rho \rangle
    \left[ \rho_0 -\langle \rho \rangle \right] - \sigma^2(\tau),\\
    \label{eq:imp2}
    \frac{d \sigma^2}{d \tau} & = 2\sigma^2\left[ \rho_0 +\langle \rho
      \rangle \right] - 2\Delta_3(\tau) +\frac{1}{N}\langle \rho
    (1-\rho)\rangle + \frac{\gamma}{N\alpha}\langle\rho\rangle,
  %% \frac{1}{2\alpha}\frac{d}{dt}\sigma^2(t) = \left[\rho(t) + \rho_{0} +\frac{1}{N}\right]\sigma^2(t)
  %% - \left[\langle \rho^3 \rangle - \rho^3(t)\right] -\left[  \frac{\rho + \rho_0}{2}-1\right]\frac{\rho(t)}{N}.
  \end{align}
  \end{subequations}
%\end{widetext}
where $\Delta_3(\tau) = \langle \rho^3(\tau)\rangle - \langle
\rho(\tau)\rangle^3$. These results find excellent agreement with
simulated data using an ensemble with $10^6$ replicas starting from
the same initial condition (see  Fig.~\ref{fig:fig1}).

%% \begin{figure}
%%   \includegraphics[width=0.75\columnwidth]{fig2.eps}
%%   \caption{\label{fig:fig2} Numerical simulations of the SIS
%%     model. Stochastic effects are far more prominent for small
%%     population sizes ($N=50$, $\gamma/\alpha = 1/2$ ), reducing the accuracy of compartmental
%%     equations. The time derivative of the density of infected $\rho$
%%     extracted directly from data (cross) using forward-derivative
%%     agrees with the RHS of Eq.~(\ref{eq:imp1}), as a function of the
%%     density and variance. The dashed line shows the expected RHS of
%%     compartmental equation Eq.~(\ref{eq:compartmental}). The equation of
%%     motion (line) for $d\sigma^2/d\tau$ in Eq.~(\ref{eq:imp2}) also agrees with
%%     simulated data (circles).}
%% \end{figure}

Comparing Eqs.~(\ref{eq:compartmental}) and (\ref{eq:imp1}), the case  
that considers temporal fluctuations decays faster than the
compartmental equation by $\sigma^2(\tau)$, even in the regime $N \gg
1$. Both equations are equivalent whenever $\sigma(\tau)$ becomes
irrelevant compared to $\langle \rho \rangle$. Therefore, a
generalization of compartmental equations for the SIS model is readily
available by retaining both mean and variance, neglecting higher
statistical moments. Thus, the dynamical system describes a gaussian
variable evolving along time. The skewness coefficient vanishes as a
direct consequence of this assumption, so that $\Delta_3(\tau) \approx
3 \langle \rho(\tau)\rangle \sigma^2(\tau).$  
%% \begin{equation}
%% \Delta_3(\tau) =\langle \rho^3(\tau)\rangle - \langle \rho(\tau)\rangle^3 \approx 3 \langle \rho(\tau)\rangle \sigma^2(\tau).
%% \end{equation}
For $N\gg 1$, the resulting equations are
\begin{subequations}
  \label{eq:system}
  \begin{align}
    \label{eq:improved1}
   \frac{d}{d\tau}\ln \langle \rho\rangle &= \rho_{0}-\langle \rho\rangle -
   \frac{\sigma^2}{ \langle \rho\rangle},\\
   \label{eq:improved2}
  \frac{1}{2}\frac{d}{d\tau}\ln\sigma^2  &=\rho_0 - 2\langle \rho\rangle.
\end{align}
\end{subequations}
We emphasize that the variance in Eq.~(\ref{eq:improved1}) slows down the
growth rate of $\langle \rho (\tau)\rangle$, recalling the Allee effect
\cite{murray,ribeiroJTheoBio2015}.

Equations~(\ref{eq:system}) can be further combined into a
single second-order differential equation \cite{nakamuraArxiv2018},
with solution 
\begin{subequations}
  \label{eq:sol}
  \begin{align}
    \label{eq:sol1}
  \langle \rho(\tau)\rangle &= \frac{{\rho_0}\left(1+c_1 \mathe^{-
      \rho_0 \tau}\right)}{1+2c_1 \mathe^{- \rho_0 \tau} + c_2
    \mathe^{-2 \rho_0 \tau}},\\
  \label{eq:sol2}
  \sigma^2(\tau) &= \frac{\langle\rho (\tau)\rangle^2(c_1^2-c_2) \mathe^{-2\rho_0 \tau}}{\left( 1 + c_1 \mathe^{- \rho_0 \tau}\right)^2}.
\end{align}
\end{subequations}
The constants $c_1$ and $c_2$ depend solely on the initial conditions.
The special case $c_2= c_1^2$ recovers the usual solution of
Eq.~(\ref{eq:compartmental}). We assumed that fluctuations behave as
gaussian fluctuations. While reasonable for various situations, the
assumption does not hold for $\gamma/\alpha $ around unity or small
population sizes, according 
to numerical simulations \cite{nakamuraArxiv2018}, in which
Eq.~(\ref{eq:imp2}) should be used instead of
Eq.~(\ref{eq:improved2}).   

%=====================================================================
%===================================================================== 
%% \section{Hamilton equations}
%% \label{sec:hamilton}
\textit{Hamilton's equations.}
The fact that the dynamical system Eq.~(\ref{eq:system}) can be combined
into a single second-order differential equation suggests an
interpretation of the epidemic model in terms of Hamilton equations
\cite{goldsteinClassicalMechanics}. Hamiltonian systems are
ubiquitous in Physics, serving as basis to describe and explain
countless physical phenomena. The hallmark of systems are the Hamilton
equations:
\begin{subequations}
  \label{eq:hami}
\begin{align}
  \frac{d q}{d \tau} & = \;\,\,\frac{\partial \mathcal{H}}{\partial p},\\
  \frac{d p}{d \tau} & = -\frac{\partial \mathcal{H}}{\partial q},
\end{align}
\end{subequations}
where $q(t)$ and $p(t)$ are conjugated variables, and the Hamiltonian
function $\mathcal{H}$ encodes some information about the problem --
usually associated with energy for conservative systems but not
restricted to them. Besides classical mechanics and related areas,
quantum field theories and statistical mechanics are deeply
intertwined with Hamilton's principle and Liouville theorem. Despite
its usefulness in Physics, Hamilton formulation and surrounding
principles are rarely used in population dynamics, ecological
problems, or epidemic models, where first-order differential equations
are dominant. The lack of second-order differential equations in these
areas, although not prohibitive, rises questions about the description
of the dynamics, as discussed extensively in
Ref.~\cite{chester2011}. In part, because it means some interactions
and forces acting on the system remains unaccounted. By adopting a
true Hamiltonian formulation, stochastic events may produce
counterintuitive effects, such as noise induced metastable states
\cite{parkerPhysRevLett2011}.

In view of the inherent stochasticity behind disease spreading and
Eqs.~(\ref{eq:system}), it seems necessary to determine whether the SIS
model is a Hamiltonian system or not. A brief inspection shows the pair
$(\langle \rho \rangle, \sigma^2)$ does not satisfy the usual Hamilton
equations. The solution to this issue is obtained by assuming,
instead, that the correct conjugated pair is $(\langle \rho \rangle,
h(\sigma^2))$, where $h(x)$ is some analytical function. Inspiration
from common pairs of conjugate variables can be used to refine the
choice of $h(x)$. For instance, the product $\langle \rho
\rangle\times h(\sigma^2)$ should be dimensionless, in close analogy
the scalar product between position and wave vectors. One possible
candidate is $h(x) = x^{-1/2}$, which entails $1/\sigma$ as the
conjugated variable to $\langle \rho \rangle$.  

Define the dynamical variables $q(\tau) = \langle \rho (\tau)\rangle$
and $p(\tau) = 1/\sigma(\tau)$ to describe the SIS model. In addition,
consider the following Hamiltonian %% \footnote{ It is also possible to
  %% express the conjugated pair using scaled variables $q'=\langle
  %% \rho(\tau)\rangle/\rho_0$ and $p'= \rho_0/\sigma(\tau)$,
  %% leading to $\mathcal{H}/\rho_0= q'p'(1-q')+(1/p')$. }  
\begin{equation}
  \label{eq:hamiltonian}
  \mathcal{H} = q(\tau)p(\tau)\left[ \rho_0 - q(\tau) \right] +
  \frac{1}{p(\tau)}.
\end{equation}
%% in which $F(\tau)$ depends explicitly on time but not on $q$ nor
%% $p$.
Plugging these expressions in Eqs.~(\ref{eq:hami}), one obtains the
equations of motion: 
\begin{subequations}
\begin{align}
  \frac{d q}{d \tau} &= q(\rho_0 - q) - \frac{1}{p^2} \equiv  \langle
  \rho \rangle[\rho_0 - \langle  \rho \rangle] - \sigma^2,\\ %% =
  %% \frac{d\langle \rho \rangle}{d\tau},\\  
  \frac{d p}{d \tau} &= -p(\rho_0 - 2 q) \equiv -\frac{1}{\sigma}[\rho_0 -
    2\langle  \rho \rangle]. %= -\frac{1}{\sigma^2}\frac{d \sigma}{d\tau}.
\end{align}
\end{subequations}
Thus, at first glance $\mathcal{H}$ appears to be a
valid candidate to describe the SIS model. Even more, replacing
$(q,p)$ by Eq.~(\ref{eq:sol}) in Eq.~(\ref{eq:hamiltonian}) shows the
Hamiltonian is a constant of motion  $\mathcal{H}^{\infty} = \rho_0
c_1(c_1^2-c_2)^{-1/2}$. The upper index in $\mathcal{H}^{\infty}$ is a
reminder that calculations take place in the absence of finite size
corrections.

%\begin{widetext}
\begin{figure} 
  \includegraphics[width=0.95\columnwidth]{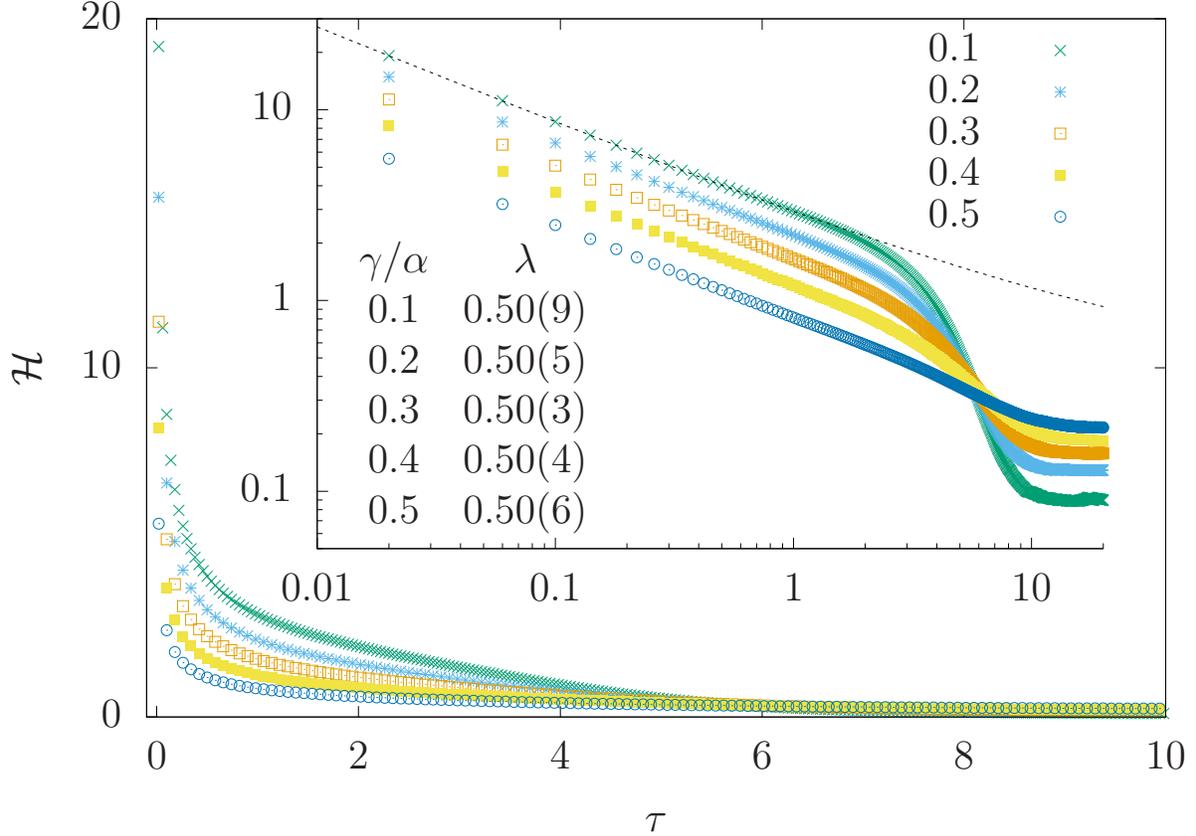}
  \caption{\label{fig:fig2} Finite size effects on the
    Hamiltonian. Simulated data with $N=50$ and $10^6$ Monte Carlo
    runs for various ratios $\gamma/ \alpha$. (inset) Initial decay of
    $\mathcal{H}$ compatible with power-law, $\mathcal{H} \sim
    \tau^{-\lambda}$. The exponent  $\lambda = 1/2$ remains constant
    for different ratios $\gamma/\alpha$, suggesting an universal
    behavior.}     
\end{figure}

However, taking finite size corrections into account changes  
drastically the notion of $\mathcal{H}$ as a constant of motion. 
In fact, as Fig.~\ref{fig:fig2} depics, numerical simulations for
finite populations reveal $\mathcal{H}$ changes continuously along
time until equilibrium sets in, akin to a non-conservative system. A
precise meaning of $\mathcal{H}$ in the epidemiological context is
still murky, at best. A detailed analysis of correlations between
changes in $\mathcal{H}$ and the spreading pattern of real outbreaks
is mandatory to understand the action-reaction analogy. In the
meantime, it is instructive to study $\mathcal{H}$ for $\tau \ll 1$
and $\tau \gg 1$ (see Fig.~\ref{fig:fig2}). For $\tau \ll 1$, where
incidentally fluctuations varies the most (see Fig.~\ref{fig:fig1}), a
remarkable feature appears via the relation $\mathcal{H}\sim
\tau^{-\lambda}$ with $\lambda = 1/2$. In particular, the exponent
$\lambda$ seems insensitive to changes in the epidemiological
parameter $\gamma$. This parameter-free behavior is not 
observed for the remaining statistics, $\langle \rho(\tau)\rangle$
 and $\sigma(\tau)$. Power-laws are  crucial to identify scaling
 relations and emergence of universal features, and they are usually
 related to the symmetry of the problem rather than microscopic
 details. Here, evidence of universal behavior is captured by the data
 collapse  $\mathcal{H}/\rho_0^2$ (not shown).
 From these observations, we can infer fluctuations play a larger role
 in the early disease spreading, being largely independent of exact
 values of epidemiological parameters.

An effective decay $\textrm{e}^{-\tau/\tau_{\textrm{eff}}}$ describes
the general behavior of $\mathcal{H}$ in the low temperature
regime. The relaxation time $\tau_{\textrm{eff}}$ depends on $N$ and
the ratio $\gamma/\alpha$, and it can be estimated from data by
fitting $\mathcal{H}$ to an exponential function plus a constant.
Alternatively, it can be evaluated as 
\begin{equation}
  \tau_{\textrm{eff}} = \frac{1}{\mathcal{H}(0)}\int_{0}^{\infty} d\tau
      [\mathcal{H}(\tau)-\mathcal{H}(\infty)].
\end{equation}
%% From a formal point of view, the evaluation of $\tau_{\textrm{eff}}$
%% requires the substitution of Eqs.~(\ref{eq:sol1}) and (\ref{eq:sol2}) in
%% Eq.~(\ref{eq:hamiltonian}), followed by an integration. The relevant input
%% data requires the initial constants $c_1$ and $c_2$, in addition to
%% $\alpha$ and $\gamma$. Surely, the procedure is arguably more
%% demanding than estimating $R_0$, as it includes an integration  beside
%% two extra parameters. However, as others have 
%% reasoned before, $R_0$ provides a naive estimation on secondary
%% infections because the growth rate of the outbreak changes
%% continuously along time \cite{feffernanJRSoc2005}.  In contrast,
%% $\tau_{\textrm{eff}}$ mimics a constant of motion.
From a formal point of view, the evaluation of $\tau_{\textrm{eff}}$
requires the solutions of Eqs.~(\ref{eq:imp1}) and
(\ref{eq:imp2}) in Eq.~(\ref{eq:hamiltonian}), followed by an
integration. Surely, the procedure is arguably more demanding than
estimating $R_0$. However, as others have reasoned before, $R_0$
provides a naive estimation on secondary infections because the growth
rate of the outbreak changes continuously along time
\cite{feffernanJRSoc2005}.  In contrast, $\tau_{\textrm{eff}}$ mimics
a constant of motion.

%=====================================================================
%===================================================================== 
%% \section{Lagrangian and canonical transformations}
%% \label{sec:lagrange}
\textit{Lagrangian.}
Another insight from $\tau_{\textrm{eff}}$ links the temporal integral
of $\mathcal{H}$ with the mechanical action $S$. A formal connection
with $S$ is desirable because it brings a large machinery revolving
around variational principles and conservation laws. However, the
action $S = \int d\tau \mathcal{L}(q,\dot{q};\tau)$ is a functional of
the Lagrangian $\mathcal{L}$. It turns out that $\mathcal{L}$ can
obtained from $\mathcal{H}$ by inspection. From Eqs.~(\ref{eq:hamiltonian}) and
(\ref{eq:improved1}), $\mathcal{H}$ takes the following form: 
$  \mathcal{H} = p\left[ q(\rho_0 - q) + p^{-2} \right]= p (d q/d\tau) + 2/p $.
%% \begin{equation}
%%   \mathcal{H} = p\left[ q(\rho_0 - q) + \frac{1}{p^2} \right]= p\frac{d
%%     q}{d\tau} + \frac{2}{p}.
%% \end{equation}
Recalling the formal expression $\mathcal{H} = p\dot{q} -
\mathcal{L}$, it becomes clear that
\begin{equation}
  \label{eq:lagrangian}
  \mathcal{L} = -\frac{2}{p} =-2 \sigma(\tau)= -2 \sqrt{q(\rho_0 - q) -\frac{d q}{d\tau}},
\end{equation}
where we have used Eq.~(\ref{eq:improved1}) and considered only the
positive root. Thus, $\mathcal{L}$ is proportional to the standard
deviation while the action entails the accumulated deviation over the
course of the outbreak. To check our result for large populations
$N\gg 1$, the minimal action recovers Eq.~(\ref{eq:compartmental}) as
expected for a noise-free system. In general, the equation of
motion reads
\begin{equation}
  \frac{d^2q}{d \tau^2} = 3 (\rho_0-2 q)\left[ \frac{d q}{d\tau} -
    \frac{2}{3}q(\rho_0-q) \right]. 
\end{equation}

The fact that $\mathcal{L}$ contains solely the standard deviation
allow us to understand how to add uncorrelated fluctuations into the model.
By virtue of $\textrm{Var}[x+y] = \textrm{Var}[x]+\textrm{Var}[y]$ for
uncorrelated random variables $x$ and $y$, the perturbed Lagrangian
can be obtained by adding a $\sigma_{\textrm{ext}}^2(\tau)$ to the
variance of the system $\sigma^2(\tau)$:
\begin{equation}
  \mathcal{L}'=-2 \sqrt{q(\rho_0 - q) -\frac{d q}{d\tau} +
    \sigma_{\textrm{ext}}^2(\tau)}.
\end{equation} 
This picture is consistent with addition of a noise function
$\sigma_{\textrm{ext}}^2(\tau)$ to Eq.~(\ref{eq:improved1}). The
perturbed Lagrangian $\mathcal{L}'$ describes, ultimately, the time
evolution of the disease prevalence in environments with noise. Note
that this description differs from the usual derivation of Langevin
equations, in which the noise function (force) $r(\tau)$ couples
linearly with $q$, i.e., $\mathcal{L}'=\mathcal{L}-r(\tau)
q(\tau)$. By the same token, the addition of correlated signals
$\eta(\tau)$ to the Lagrangian entails corrections from the  covariance
matrix: since $\textrm{Var}[X+Y]=\textrm{Var}[X]+\textrm{Var}[Y]+2\textrm{Cov}[X,Y]$,
then $\mathcal{L}'=-2 \sqrt{\sigma_{\rho}^2 +\sigma_{\eta}^2 + 2
  \textrm{Cov}[\rho,\eta]}$. The covariance matrix can estimated or
modeled directly from data, promoting further understanding on the
spreading of co-existing diseases, where facilitation or competition
processes are in place.  

With both Hamiltonian and Lagrangian formalisms secured, canonical
transformations become available. These transformations are
particularly useful to highlight properties of the dynamical systems
and to solve them. They change the old variables $(q,p)$ into new
variables $(Q,P)$, while preserving Hamilton's equations. There are a 
large number of transformation available: it would render impossible
to cover all of them here. Instead, we show that at least one
canonical transformation exists, and that it promotes the
interpretation of the stochastic spreading process as effective
mechanical systems.  Consider: $P_1(t) = 2 p^{1/2} q $ and $Q_1(t) = -
p^{1/2}$. The Poisson bracket $\{Q_1,P_1 \}_{q,p} = (\partial
Q_1/\partial q)(\partial P_1/\partial p) - (\partial Q_1/\partial
p)(\partial P_1/\partial q) = 1$ shows the transformation is
canonical. Setting  $m=2$, the Hamiltonian in terms of the canonical
variables $(Q_1,P_1)$ becomes  
\begin{equation}
  -\mathcal{H}_1 = \frac{1}{2m} \left(P_1+\rho_0 Q_1 \right)^2 -
  \frac{\rho_0^2Q_1^2}{2m} - \frac{1}{Q_1^2}. 
\end{equation}
One may interpret $-\mathcal{H}_1$ as the Hamiltonian of an effective
mechanical problem in one-dimension, in which the particle has mechanical 
momentum $P_1(\tau)$, with generalized coordinate $Q_1(\tau)$,
subjected to a velocity dependent potential.

%=====================================================================
%===================================================================== 
%% \section{Conclusion}
%% \label{sec:conclusion}
\textit{Conclusion.}
The description of several real world problems often contains
stochastic fluctuations. The SIS epidemic model includes them due to
uncertainties associated with pathogen transmission. For small
fluctuation amplitudes, $\langle \rho(\tau)\rangle$ and
$\sigma^2(\tau)$ are adequate descriptors. Our findings demonstrate
$\langle \rho(\tau)\rangle$ and $1/\sigma(\tau)$ are conjugated
variables, and they satisfy Hamilton's equation. These results link
the stochastic SIS epidemic model with a pure dynamical system, which  
can be solved and manipulated using standard analytical tools.
We find the Hamiltonian is a constant of motion for $N\gg 1$. However,
finite size effects break the temporal symmetry of the system:
$\mathcal{H}\sim \tau^{-1/2}$ follows a  power-law around the
outbreak onset. A clear explanation for this scaling is still
lacking. The relaxation time $\tau_{\textrm{eff}}$ portrays the decay
of $\mathcal{H}$ until equilibrium sets in, meaning that it can also
be used to characterize the SIS epidemic. Unlike popular estimates of
epidemic growth rate, such as $R_0$, $\tau_{\textrm{eff}}$ remains
constant along time and can be extracted from data values of
$\mathcal{H}$. Finally, our results also suggests a way to incorporate
interactions into the SIS model via the Lagrangian function. This
finding has intriguing implications for our understanding of
facilitation-competition mechanisms between co-occurring diseases
since it does not replicate the canonical procedure to obtain Langevin
equations.

\begin{acknowledgments}
  The authors acknowledge funding CNPq 307948/2014-5 and Capes
  88887.136416/2017-00. 
\end{acknowledgments}

%=====================================================================
%===================================================================== 
%\bibliography{bibdatabase}

%merlin.mbs apsrev4-1.bst 2010-07-25 4.21a (PWD, AO, DPC) hacked
%Control: key (0)
%Control: author (0) dotless jnrlst
%Control: editor formatted (1) identically to author
%Control: production of article title (0) allowed
%Control: page (1) range
%Control: year (0) verbatim
%Control: production of eprint (0) enabled
%

\end{document}